# Anomaly Detection of Smart Metering System for Power Management with Battery Storage System/Electric Vehicle

Sangkeum Lee[1] | Sarvar Hussain Nengroo[2] | Hojun Jin[2] |Yoonmee Doh[1] | Chungho Lee[1] | Taewook Heo[1]| Dongsoo Har[*2]

[1] Environment ICT Research Section, Electronics and Telecommunications Research Institute (ETRI), South Korea, Daejeon
[2] Cho Chun Shik Graduate School of Mobility, Korea Advanced Institute of Science and Technology (KAIST), South Korea, Daejeon.

**Correspondence**
*Dongsoo Har, Cho Chun Shik Graduate School of Mobility, Korea Advanced Institute of Science and Technology (KAIST), South Korea, Daejeon. Email: dshar@kaist.ac.kr

**Funding Information**
This work was supported by the Korea Institute of Energy Technology Evaluation and Planning (KETEP) and the Ministry of Trade, Industry & Energy (MOTIE) of the Republic of Korea (No.20191210301580, No. 20192010107290).

**Abstract**

A novel smart metering technique capable of anomaly detection was proposed for real-time home power management system. Smart meter data generated in real-time was obtained from 900 households of single apartments. To detect outliers and missing values in smart meter data, a deep learning model, the autoencoder, consisting of a graph convolutional network and bidirectional long short-term memory network, was applied to the smart metering technique. Power management based on the smart metering technique was performed by multi-objective optimization in the presence of a battery storage system and an electric vehicle. The results of the power management employing the proposed smart metering technique indicate a reduction in electricity cost and amount of power supplied by the grid compared to the results of power management without anomaly detection.

## 1 | INTRODUCTION

Air pollution, climate change, and depletion of fossil fuel resources are key societal issues in this century. The transportation and electric power generation industries, which are among the largest consumers of fossil fuels, have voiced these concerns. In the United States, the industrial sector is the largest energy consumer [1], consuming 31% of the total energy and accounting for approximately one-third of total greenhouse gas (GHG) emissions in the country [2,3]. In the United Kingdom, households accounted for 39% of the total electricity consumption and 9% of the carbon dioxide ($CO_2$) emission in 2019 [4]. Therefore, households have great potential to realize the net-zero emission goal by implementing solutions like decentralized electricity production at home and reducing electricity use [5].

When considering the photovoltaic (PV) power battery system's storage component, the storage system increases the local generating self-consumption while decreasing power costs, fossil fuel generation, and the strain on the electricity distribution infrastructure [6]. The main reasons why households should embrace renewable energy are economic and environmental benefits [7]. The US government established a goal of reducing energy consumption by 17%, below 2005 levels, by 2020. As a result, technologies that promise to cut GHG emissions while deferring or avoiding large additional investments have piqued public attention. The energy efficiency improvements combined with demand response (DR) are expected to lower the need for new generation capacity from 214 to 133 GW in 2030 [8]. Additionally, the DR has been proposed as a strategy to address supply-demand oscillations in the grid with significant penetration of variable renewable energy sources (RESs) of intermittent nature [9]. With the high penetration of RESs, DR has also been advocated as a solution to control the supply-demand oscillations in the grid [10,11]. Several research findings have provided customers with information in various areas, allowing them to better regulate the amount of electricity used by interruptible, non-interruptible, shiftable, and non-shiftable devices in reaction to the changing price. Human and automatic control systems have been extensively studied to lower the electricity demand of buildings during peak periods [12,13].

The complexity of the power system has been increasing, particularly in the distribution grids, and suitable measurement infrastructure should be deployed [14,15]. To deal with these challenges and achieve the modernization of the conventional electricity infrastructure, smart metering and control systems that operate based on the interaction between suppliers and consumers are essential. A smart metering infrastructure (SMI) is described as an electronic system that can measure energy consumption via smart meters (SMs) by providing more diverse information than the existing infrastructure and can send/receive the usage details directly to/from other parties through electronic communication networks. When considering the scope of advanced metering infrastructure (AMI) innovations, SMs have emerged as an incredible asset, particularly for data analytics. Due to its large volume, velocity, and variety, SM data have significantly enhanced the degree of information gathering across dispersed networks and have acquired fame in big data analytics. Furthermore, the large volume of SM data allows optimal real-time monitoring and control of electric utilities to increase reliability and operational efficiency, as well as enhance key performance measurements, including system average interruption duration index and customer average interruption duration index.

As the electricity market is based on a cost mandatory pool system in Korea, the fuel cost is considered a significant component. The market price consists of the capacity payment (CP) and system marginal price (SMP). The CP describes the price provided to a generating unit available for a day, and the SMP represents the cost of the most expensive generating unit obtained in the price setting schedule, which minimizes the total operational costs of the generating units. Additionally, it depends on the bidding price of a marginal plant predetermined by the fuel cost and plant performance. Energy prosumers with PV system contracts with Korea Electric Power Corporation can obtain a profit from the power purchase agreement by selling surplus electricity at SMP, which is the most common electricity trading method for energy self-consumption [16]. Furthermore, through the electricity trading system, surplus electricity can be directly traded to the nearby energy consumers or prosumers as a form of peer-to-peer (P2P) transaction [17].

Measuring and verifying the energy-saving effects is essential when an operational strategy is developed with an investment to increase energy efficiency. After implementing energy conservation measures, the measure and verification (M&V) process is used to plan, measure, collect, and analyze data, which is applied to verify and report the energy-saving effects of various facilities. To increase energy efficiency, the international performance measurement and verification protocol (IPMVP) was developed by the efficiency valuation organization. To increase the certainty and reliability of the energy-saving effects and reduce transaction costs with an M&V plan, IPMVP provides the framework and evaluation criteria for M&V, and IPMVP is a globally recognized protocol. The measured data are integrated with the process of developing, installing, and operating energy-saving strategies through routine and non-routine adjustments. The routine adjustments are applied to the overall energy management factors, which are expected to change periodically, such as weather and production volume. Routine adjustment can be achieved using several techniques, including constant value (no adjustment) or multiple parameter non-linear equations with independent variables. The non-routine adjustments are applied to energy management factors that are not expected to change, such as the size and design of the facility, the operation of installed equipment, and the type of occupant. However, these factors should be monitored in order to prepare for changes during the operating period. Therefore, the M&V plan calculates saving effects by comparing the measured consumption before and after the implementation of an operational strategy, making appropriate adjustments for changes, such as occupancy, production, and weather.

The growing use of information and communication innovation in smart cities is causing a major shift in the energy management paradigm for power networks and buildings [18]. Building energy management is a critical responsibility for improving energy efficiency and reducing the mismatch between actual and expected energy consumption, which is frequently caused by erroneous occupant behavior or equipment and control system failures [10]. To increase energy efficiency, comfort, and equipment life, while also lowering energy consumption and operating expenses, it is crucial to have an efficient management system. Anomaly detection is incredibly helpful in enhancing the performance of building energy management, and it is promising in terms of cost reduction when incorporated into the energy data detection strategy [11]. Clustering algorithms and other machine learning algorithms with long short-term memory (LSTM) networks have been applied to SM data in recent studies, mainly in the context of anomaly detection [19,20]. There is an immense potential for anomaly detection at various levels of research in the area of building energy management. This sort of investigation has been primarily concentrated at the framework level (e.g., heating, ventilation, and air conditioning (HVAC) systems). Furthermore, in a liberalized energy market, energy profiling is highly desirable for load prediction [12,13], tariff setup description [14], client organization [21], employing targeted demand side management strategy [15], encouraging building energy demand changes, and executing DR schemes.

The following are the main contributions of this work:
1. A novel smart metering technique capable of anomaly detection is proposed for real-time home power management. The proposed technique for anomaly detection can detect outliers and missing values in the SM data. In fact, the smart metering technique also enables the prediction of new SM data.
2. A predictive type of power management is implemented with the assistance of anomaly detection to reduce peak load and electricity costs.
3. A deep neural network (DNN) model, the autoencoder, consisting of a graph convolutional network (GCN) and bidirectional long short-term memory (BiLSTM) network, is adopted for the anomaly detection and the prediction of SM data. Predicting the future value of SM data enables power management to be more cost effective.
4. Optimization with a multi-objective function is proposed to increase the profits of renewable energy providers (REPs). The REPs may have reactive clients due to the lack of a home energy management system (HEMS) or a customer's unwillingness to participate in DR programs. These clients tend to choose the energy usage or manage it based on their expertise or experience rather than evaluating the best plan [8]. The REPs may not be aware of the function of the reactive customers' reaction pattern in these circumstances. Therefore, the REPs must also learn about the consumers' energy usage patterns.

The remainder of this article is organized as follows. In Section 2, a review of the home power management is presented. Section 3 presents an overview of reported studies dealing with smart metering systems. Section 4 provides a description of the cloud-based smart metering system. Section 5 presents the DNN model for anomaly detection and prediction of SM data, as well as power management based on the DNN model. Section 6 is for the discussion of the results of power management. Section 7 concludes this paper.

## 2 | LITERATURE REVIEW

Presently, the applications of renewable energy power transformation are evidently accessible both technologically [22] and monetarily [23]. In the Kuda Bandos Island of Maldives, Jung et al. [24] assessed three potential system configurations using HOMER® software to determine which configuration would result in the most optimal off-grid energy management. Their findings indicate that a PV system can be a cost effective alternative for the resort, and grid parity can be achieved within the project's lifetime. The levelized cost of electricity and the time taken to reach grid parity when solar PV and battery storage are deployed were analyzed in their study. Recently, studies dealing with smart grids operated with the DR system have been conducted, regarding energy management with optimization techniques, such as mixed-integer linear programming [25], geometric programming [26], model predictive control [27], dynamic programming [28], and stochastic dynamic programming [29].

Electric utility companies all around the world have been focusing on power system stability and blackout management. In recent years, smart grids have contributed to boosting dependability, versatility, and stability, as well as enhancing energy management across power utilities by coordinating the needs and capacities of generation sources, grid machinists, and end-users through remote communication. According to a recent study [30], when individuals' work profiles are integrated into building energy management systems, 10%–40% of electricity can be saved in households. Capozzoli et al. [31] proposed an overlay structure for the mining of regular load profiles at solitary and different structure levels and explored several applications connected to the analysis of energy profiles in buildings. Particular information analytics methods, such as clustering algorithms, have been used to address energy profiling in structures [32,33]. G. Chicco [32] provided an overview of the various clustering approaches and clustering validity pointers used to assemble similar load profiles. Fernandes et al. [33] used the fuzzy C-means method to investigate one year-long hourly gas usage information of more than 1,000 buildings to derive typical patterns linked with the respective consumer groups and discovered that the peaks of the morning and evening usage primarily defined the agent profiles.

Developing an intelligent HEMS has become an important goal to support the trend toward a more sustainable energy supply for the SG. One of the key aspects of the SG is the intelligent HEMS, which automatically adjusts household loads through a link between consumers with options of SMs, smart appliances, electric vehicles, and home power generation and storage systems. Based on the real-time supply and demand of microgrids, a distributed dynamic pricing strategy for plug-in hybrid electric vehicle management integrated with smart grids was devised in [34]. The optimal portfolio selection technique was used by the authors in [35] to build a dynamic appliance scheduling scheme. Additionally, metaheuristic optimization was used to optimize the scheduling of shiftable loads in an SG [36]. A Stackelberg game in [37] was used to arrange the optimal regulation of home appliances via virtual power trading. A deterministic DR model for energy usage optimizer software, such as HEMS, in which the day-ahead cost of electricity is known ahead of time, was discussed in [38,39]. A benchmark has been presented in which the HEMS combines an artificial neural network to anticipate future pricing and multiagent reinforcement learning to select the best options for household appliance.

## 3 | RELATED WORKS - SMART METERING SYSTEM

In the past few years, smart metering systems have received significant attention from European Union (EU) countries. These systems have been significantly influenced by the EU Energy Efficiency Directive, aiming to minimize the negative environmental impacts of energy generating units and satisfy the commitments relevant to climate change under the Kyoto protocol [40]. Also, the structure of the smart metering system is standardized by the European Federation cooperating with the SM coordination group CEN-CENELEC-ETSI. In the standardized infrastructure, the distribution system operator (DSO) manages the grid operation in the designated region and is responsible for installing and maintaining SMs in each household. The energy suppliers are commercial groups that generate or purchase electricity from the self-energy production system or electric utility and sell it to customers, using the infrastructure of the DSO to deliver this electricity. Additionally, the infrastructure provides smart service based on the shared data by the energy suppliers, energy traders, and energy service companies while prioritizing customer privacy protection. With the introduction of SMs, the independent service providers (ISPs) can offer optional services, such as providing a detailed analysis of electricity usage and guidelines to save energy consumption.

The Netherlands has deployed a smart metering system for the home area, which measures electricity usage and delivers information related to gas, heat, and water meters to a data collection server by connecting meters to an integrated interface. The energy supplier obtains the customer's meter readings provided by the DSO to issue the bill to the customer. The Dutch DSOs established the energy data services Nederland as a central organization to facilitate information delivery management instead of supervising the many-to-many relationships between the DSOs and energy suppliers. Additionally, the energy data services Nederland has an obligation to provide SM data to the energy suppliers and ISPs, superseding the right of the DSO in each region. The specification of SMs is regulated by the Dutch SM requirements and standards [41]. As most households have a connection with natural gas, SMs simultaneously access the readings of the electricity and gas. Besides measuring energy consumption, SMs also measure the power quality and outage by providing time synchronization and shifting between tariffs.

The German smart metering strategy, which places a strong emphasis on standardization and security, is built on two primary components: SMs and smart meter gateways, and the combination of the two is referred to as a smart metering system. In Germany, a significant share is achieved in renewable energy due to guaranteed feed-in tariffs of renewable energy and well-conditioned grid infrastructure. Market-oriented tariffs and higher intelligence are required for power grids to meet rapidly increasing renewable energy demands. Therefore, the SG applications, such as the provision of secondary energy resources by virtual power plants, feed-in management through direct control of distributed energy resources (DERs), DR management, and applications to change the consumers' behavior, are required to be addressed by German regulation and the grid operator. Secondary reserve by the DER and feed-in management has been previously established, and the DR program management to influence customer behavior is being addressed. In Germany, several research and development projects, such as the decentralized energy management systems [42] and virtual power plants [43], the provision of decentralized ancillary services from renewable energy generation units, have been initiated [44]. Detailed metering data and reliable communication with the DER and consumers are required to improve the performance of these SG applications. The bill "digitalization of the energy transition" mainly deals with the contents of infrastructure regulation, security problems, and price limitations. As the SMI of Germany is implemented, it is required to provide DSOs, energy suppliers, consumers, and authorized groups with smart metering and communication services, such as collection and storage of SM data depending on German regulations, and the provision of secure and private protection of the SM data. In the regulation of the German SMI, a few key roles and components are differentiated. In the system configuration, there are a digital SM, an SMGW, and a controllable local system (CLS). The SMGW is described as a communicable device, which collects information from the SMs and delivers it to the assigned receivers. Moreover, it is used to establish reliable communication channels to control local CLS devices, achieving security and privacy requirements. The SMGW administrator has an independent authority to check authentication and qualification for the certification of the authorized external entity (AEE), which manages the demand for SM data or communication with CLS devices located on the customer's site. Depending on the passive and active AEE, regulations of smart metering systems are differentiated. While the passive AEE only receives the SM data from the SMGW, the active AEE can interact with CLS devices.

## 4 | CLOUD-BASED SMART METERING SYSTEM

### 4.1 | Structure of smart metering system in the home area

The SMs monitor, measure, and deliver the real-time energy consumption data to utility providers based on the user activity to analyze and control power network systems. Compared with the traditional meter, the SMs provide more sophisticated services, such as two-way real-time communication between utility providers and SMs, time-based demand data analysis, measuring service quality, outage management, distribution network analysis and planning, customer billing, demand reduction, and remote connection. Additionally, the SM enables continuous reading and recording of the gas, electric, water, heating, and hot water utilities. The general architecture of the smart metering system consists of some units that provide the required amenities to the consumers. Fig. 1 depicts the overall framework of the smart metering system.

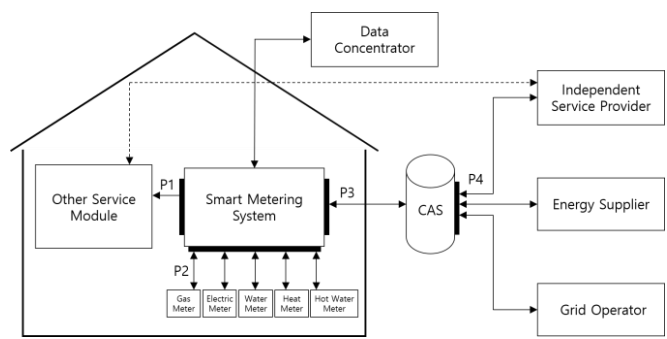

**FIGURE 1** Overall architecture of the smart metering system.

A data concentrator is described as a device typically installed in substations that manages metering data acquired from multiple SMs in distant households. Delivering data packets from SM to the data concentrator can be configured in wireless communication networks, such as orthogonal frequency-division multiplexing [45,46]. Also, the data concentrator primarily acts as a store-and-forward connection between SMs and the rest of the system, gathering data on energy usage at remote residences, relaying the data to the control center, and delivering the data to the billing system. Furthermore, if required, it can find and arrange newly installed SMs while generating repeating chains. Generally, a portion of infrastructure can be managed autonomously by the data concentrator, such as keeping track of the power grid and SMs, affirming malfunctions and disturbances, and detecting and probing obstructing efforts. While the data concentrator cannot accept incoming calls, it can initiate and continue communication if the link is lost. As described in NTA 8130, the central access server (CAS) is used as the central application to administer information collection, control, and parameterization.

Additionally, it serves as a centralized authorization system to provide authority for accessing the metering system. As part of the metering framework activity, every grid operator keeps a collection of servers, and these servers feature various programs and applications that are essential to the system's operations. Moreover, these servers act as Internet gateways for clients, allowing them to access their profiles, and monitor their energy consumption patterns. Also, the metering dataset provided by the servers can be used as the input and output of the autoencoder for training purposes. Metering data are used by the ISPs to provide non-essential services, such as saving electricity through service modules. Commercial entities that generate or purchase electricity and sell it to customers are referred to as energy suppliers, and when allocating electricity usage, grid managers often use the information provided by the metering system.

The smart metering system has multiple ports through which information can pass to facilitate communication between system components and market participants. As shown in Fig. 1, four ports are specified by the NTA 8130 standard: P1, P2, P3, and P4. The read-only port P1 is used to link the metering installation to external devices. Other local metering equipment can be connected to port P2, which is used to link the gas/electricity/water/heat/hot water meters via wired or wireless connections. Port P3 is used to convey information, including metering values, status, power quality, and outage measures, to the DSO. Through the long-term evolution, code-division multiple access, or packet radio service, communication between port P3 and DSO is established with the communication protocol based on the international standard IEC 62056. Port P4 is used as a gateway for ISPs, energy suppliers, and grid operators to obtain measurements from port P3. Web service to access the CAS can be provided through port P4, allowing ISPs and energy suppliers to obtain metering data from clients, regardless of the responsible DSO.

Customers can benefit from various services provided by smart metering systems, including the generation of remotely readable meter data, the facilitation of energy savings for consumers, and the monitoring of distribution networks. Additionally, the functionality of a smart metering system requires a two-way communication network. Grid operators, energy suppliers, service providers, and customers communicate via messages, providing information about the state of meter installation and the operating environment. For instance, energy suppliers and grid operators can use port P3 to display current state information on the metering system.

## 4.2 | Design of the cloud-based smart metering system

A cloud-based smart metering system is shown in Fig. 2. The data concentrator unit typically collects data relevant to energy consumption and other customers' information from the SMs of EV and PV operators through wireless or power line communication networks and transmits this data to the cloud system in real-time [47]. This transmitted data can be stored in the big data server and is verified to manage SM data in connection with the metering data check server. Based on the meter data, different application services installed in SMGW are provided to the cloud system via the advanced metering service server. Furthermore, management services, such as the HEMS, are applied to the power system through the application service server, depending on the DR program. Customers can access their meter data through the user portal, which is available in the cloud system. The proposed cloud-based smart metering system can address issues, such as the server's vast capacity and simultaneous loads on the server, by distributing cloud systems in each area and providing meter data in conjunction with the central cloud server.

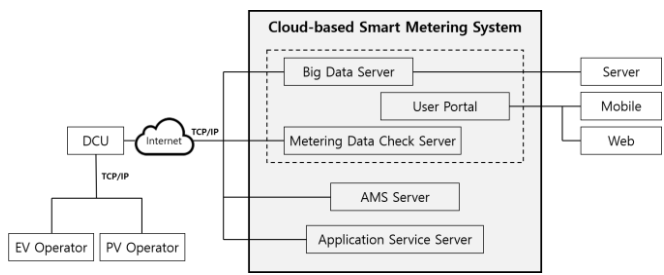

**FIGURE 2** Architecture of cloud-based smart metering system.

A brief system for the proof of concept implementation is constructed based on a structure designed to prevalidate the structure and service utility of the target metering cloud system. SM data of electric consumption collected from 900 households of single dwelling units, such as apartments, is generated in real-time through a data concentrator unit emulator, and the process of efficiently collecting, storing, processing, and visualizing the SM data is identified.

## 5 | POWER MANAGEMENT BASED ON ANOMALY DETECTION

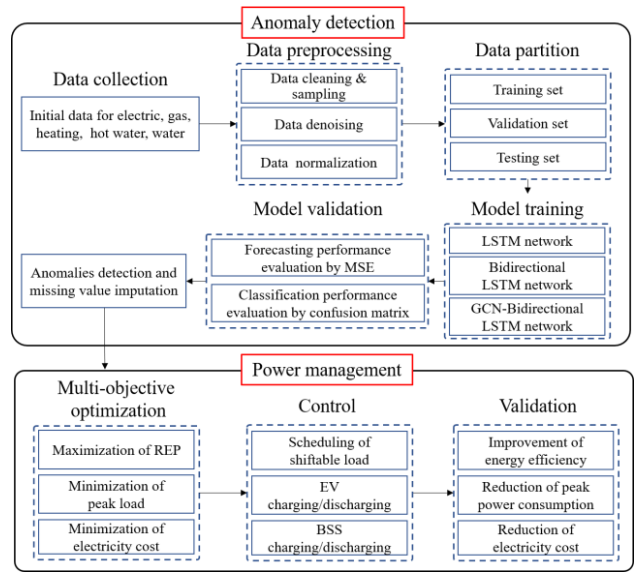

**FIGURE 3** Power management procedure with anomaly detection in a smart metering cloud system.

This section presents the proposed anomaly detection and subsequent power management. Figure 3 shows the power management procedure. The procedure for the proposed power management consists of data collection, data preprocessing, data partition, model training, model validation, and finding anomalies in the anomaly detection and multi-objective optimization, control, and validation in power management. Anomaly detection can be actively and efficiently applied to the power management scheme with BSS/EV charging/discharging methods by removing outliers and missing values in SM data. In the data collection step, the electricity consumption data for energy resources are collected from the smart metering system. Data cleaning, sampling, denoising, and normalization processes are executed in the data preprocessing step to reduce the complexity obtained in the collected real-world dataset. To train and evaluate the performance of the anomaly detection model, the collected dataset is divided into a training set, validation set, and testing set in the data partition step. In the model training and validation step, the anomaly detection model is trained with a training set, and the forecasting performance and classification performance of the model are evaluated by mean square error (MSE) and confusion matrix, respectively. Based on the results of anomaly detection, power management is performed. In the multi-objective optimization and control step, the charging/discharging operations of BSS/EV are determined depending on the specific condition, and the scheduling of shiftable loads is executed through multi-objective optimization, which aims to maximize REP and minimize peak load and electricity cost. In the validation step, the proposed power management method is evaluated in terms of improvement in energy efficiency and reduction in peak power consumption and electricity cost.

## 5.1 | Anomaly detection

In the smart metering system, outliers and missing values can randomly occur in nature and are typically caused by a metering device not delivering measured values or by a faulty measuring instrument [48]. Moreover, outliers and missing data are caused by failure of measuring instruments, the poor performance of measuring sensors, errors during installation and poor protection, intentional damage, power outages, flicker, and phase loss. These values can incur a gap or discontinuity, and the omission of detecting these anomalies reduces the accuracy and energy waste in the power management process. Therefore, it is essential to identify outliers and missing values and remove these values before applying a power management scheme. The deep learning model is established using an autoencoder to detect outliers and missing values for anomaly detection. Unsupervised learning, which is a method for predicting results for new data by clustering data without labels of correct answers, is used for the autoencoder.

### 5.1.1 | GCN-Bidirectional LSTM network

Convolution in GCNs is basically multiplying the input neurons with a set of weights, which are commonly known as filters or kernels. The same filter will be used throughout the image within the same layer, referred to as weight sharing. GCNs perform similar operations in which the model learns the features by inspecting neighboring nodes. The major difference between CNNs and GCNs is that CNNs are specially built to operate on regularly structured data, while GCNs are the generalized version of CNNs where the number of node connections varies, and the nodes are unordered. Spatial and spectral GCNs are generally used for time series image data [49].

The LSTM network, a variant of the recurrent neural network (RNN), is particularly implemented to process time series data [50,51]. Compared to the RNN network autoencoder, the LSTM network autoencoder is generally robust for short- and long-time series data. The LSTM cell has forget, input, and output gates for short and long-term cell memory, as shown in Fig. 4. These gates regulate the interactions between the various memory units. The input gate, particularly, determines whether the input signal can adjust the conditions of the memory cell or not. The output gate determines whether it can alter the conditions of other memory cells or not. The forget gate has the option of forgetting (or remembering) its previous condition. For each element in the input sequence, each layer computes the functions as follows:

$$i_t = \sigma(W_{ii}x_t + b_{ii} + W_{hi}h_{(t-1)} + b_{hi}) \quad (1)$$

$$f_t = \sigma(W_{if}x_t + b_{if} + W_{hf}h_{(t-1)} + b_{hf}) \quad (2)$$

$$g_t = \tanh(W_{ig}x_t + b_{ig} + W_{hg}h_{(t-1)} + b_{hg}) \quad (3)$$

$$o_t = \sigma(W_{io}x_t + b_{io} + W_{ho}h_{(t-1)} + b_{ho}) \quad (4)$$

$$c_t = f_t * c_{(t-1)} + i_t * g_t \quad (5)$$

$$h_t = o_t * \tanh(c_t) \quad (6)$$

where $(W_{ii}, W_{if}, W_{ig}, W_{io})$, $(W_{hi}, W_{hf}, W_{hg}, W_{ho})$, and $(b_{hi}, b_{hf}, b_{hg}, b_{ho})$ are input weights, recurrent weights, and recurrent biases. $h_t$ is the hidden state at time $t$, $c_t$ is the cell state at time $t$, $x_t$ is the input at time $t$, $h_{(t-1)}$ and $c_{(t-1)}$ are the hidden state and cell state of the layer at time ($t$ -1) or the initial hidden state at time 0, and $i_t$, $f_t$, $g_t$, and $o_t$ are the input, forget, cell, and output gates, respectively. $\sigma$ is the sigmoid function, introduced as a logistic function, and "$*$" is the Hadamard product. The input gate determines whether new data are stored in the LSTM cell or not. Two layers exist in this gate, namely, the sigmoid and tanh layers. The sigmoid layer decides which value should be updated, while the tanh layer creates a vector of new candidate values to be stored in the LSTM cell. Equations (1)–(3) are used to determine the output of these layers. The forget gate uses the sigmoid function to determine whether data should be disconnected from the LSTM cell. The output of this gate has a value between 0 and 1, with 0 indicating that the given value should be completely discarded and 1 indicating that the entire value should be preserved. The output gate uses a sigmoid to identify which part of the LSTM is assigned to the output and provides a value between 0 and 1 by performing a non-linear tanh function.

Consequently, the output of the sigmoid layer is multiplied by the obtained result, and the output is obtained from equations (4) and (6). Two independent LSTM networks are used in BiLSTM topologies, using the first network computing data in the conventional forward sequential order and the second network computing data in the opposite order, as shown in Fig. 4. At each timestep, the output values from the forward and backward cells are linked to produce a solitary output. Forward and backward LSTM networks are initiated with identical hidden and cell states. Fig. 4 depicts the autoencoder obtained from the GCN combined with the BiLSTM network. The input of the GCN-BiLSTM autoencoder consists of reading the electric consumption, and the output is the reconstructed input data calculated by the autoencoder. Anomaly detection is implemented by comparing the difference between the input and output values with a threshold.

Individual LSTM cells can learn the context from future information by producing a reversed replica of the input data. As a result, the network can process both past and future information at any time, unlike the unidirectional LSTM network that can only process past information. Furthermore, since it is computationally affordable, the BiLSTM network uses the same backpropagation through time training approach as LSTM networks.

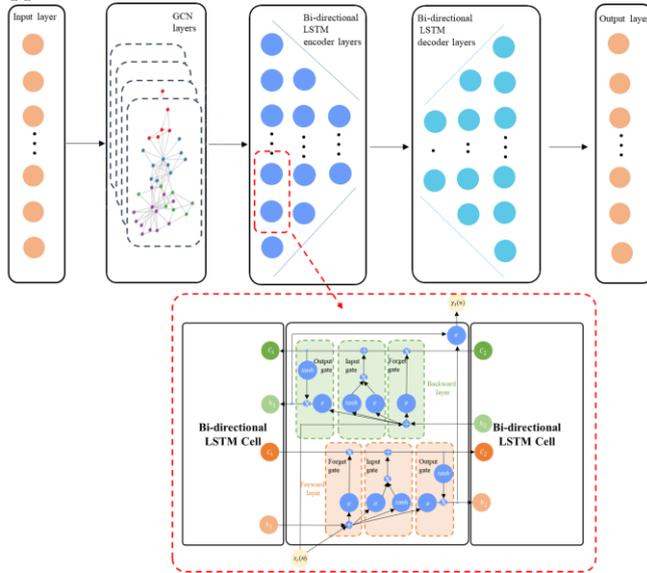

**FIGURE 4** Autoencoder implemented with GCN-BiLSTM network for anomaly detection.

### 5.1.2 | Autoencoder

A bidirectional autoencoder is an autoencoder that combines an encoder and decoder, each employing the GCN-BiLSTM network for processing the data sequence. The input sequence is entered sequentially, and after the last input sequence, the decoder regenerates the input sequence or outputs a prediction of the target sequence. For the autoencoder training, normal data without anomalies are used, and the output is calculated to measure the reconstruction error by comparing the output sequence with the input sequence. Autoencoders are trained to minimize the reconstruction error, which can be defined as the squared error as follows:

$$e_t = \|x_t - \hat{x}_t\|^2$$
$$= \|x_t - \sigma(W_{io}x_t + b_{io} + W_{ho}h_{(t-1)} + b_{ho})\|^2 \quad (7)$$

$$O_{LSTM}(t) = \begin{cases} +1 \text{ (normal data)} & e_t \leq \theta \\ 0 \text{ (outlier)} & e_t > \theta \end{cases} \quad (8)$$

where $e_t$ is reconstruction error, $O_{LSTM}(t)$ is the abnormal detection parameter at time $t$, $t$ is the 15-minute interval, and $\theta$ is the threshold for reconstruction error. The reconstruction error $e_t$ can be used to determine whether the input is anomaly data. The input is classified as normal data when the $e_t$ is lower than the threshold $\theta$, and outlier when the $e_t$ is higher than the threshold $\theta$. The $x_t$ is the input data of electric consumption, and $\hat{x}_t$ is the data reconstructed by the decoder. When an outlier or a missing value occurs in the SM data, the outlier or missing value can be replaced by $\hat{x}_t$.

In the data preprocessing stage, data cleaning and sampling, data denoising, and data normalization are used to train the GCN-BiLSTM network. When considering data cleaning and sampling, the classification and cleaning of five types of energy resource data for each household, cleaning of tag values, and resampling at 15-minute intervals are implemented. The discrete wavelet transform algorithm is applied to remove the noise, which is the cause of the degraded training performance of the GCN-BiLSTM network. Additionally, data normalization with standardization achieves the best training results for the GCN-BiLSTM network.

## 5.2 | Power management

There are numerous customers with varying energy usage behavior that can be used to cluster the grid's energy consumption. In this study, consumers on the grid may be reactive customers who do not participate in DR programs and consume energy on an as-needed basis or proactive customers who use HEMS to discover the cheapest times to use their equipment. This section shows the mathematical representation of the consumer optimization model. Power management with the smart metering system is implemented through multi-objective optimization with two objective functions. The first objective function is to maximize the profit of REPs while reducing grid dependency and increasing RES utilization. The second objective function is for customers to minimize the peak load while considering the three types of DR programs. Furthermore, the charging/discharging operations of BSS/EV are determined based on the results of anomaly detection using the GCN-BiLSTM autoencoder.

### 5.2.1 | Objective function of REP's profit maximization by minimizing grid dependency

The first objective function is proposed to lessen the grid power consumption and increase PV power utilization, aiming to maximize the profit of REPs. The power consumption of the charging/discharging of BSS/EV is also considered for power management of the smart metering system. Power management implemented for the maximization of REP's profit is advantageous in realizing the profits and yields a platform that is profitable via the SMGW. Additionally, it offers greater benefits to consumers and allows them to actively participate in the energy market.

$$\min_{\substack{(PW_{grid}(t), O_{grid}(t), PW_{PV}(t), O_{PV}(t) \\ PW_{BSS}(t), O_{BSS}(t), PW_{EV}(t), O_{EV}(t))}} \begin{bmatrix} PW_{grid}(t) * O_{grid}(t) \\ -PW_{PV}(t) * O_{PV}(t) \\ +PW_{BSS}(t) * O_{BSS}(t) \\ +PW_{EV}(t) * O_{EV}(t) \end{bmatrix} \quad (9)$$

where $PW_{grid}(t)$ is the grid power consumption for load demand at time $t$, $O_{grid}(t)$ is the switch function for the grid power consumption at time $t$, $PW_{PV}(t)$ is the PV power consumption for load demand at time $t$, $O_{PV}(t)$ is the switch function for the PV power consumption at time $t$, $PW_{BSS}(t)$ is the power consumption for BSS charging/discharging at time $t$, $O_{BSS}(t)$ is the switch function for BSS charging/discharging at time $t$, $PW_{EV}(t)$ is the power consumption for EV charging/discharging at time $t$, and $O_{EV}(t)$ is the switch function for EV charging/discharging at time $t$. The switch functions, such as $O_{grid}(t)$, $O_{PV}(t)$, $O_{BSS}(t)$, and $O_{EV}(t)$, are controlled by the first objective function on the supply side. $O_{BSS}(t)$ and $O_{EV}(t)$ can be determined using two methods: the conventional method in equation (11) and the proposed method in equation (12).

### 5.2.2 | Objective function for minimizing peak load with DR program

A REP operates as a broker between the wholesale electricity market and consumers in the energy market. The profit of REPs is calculated by subtracting the income from selling energy and the cost of purchasing energy from the wholesale market or the cost of providing services. The cost function of the REP must be used to estimate an optimal retail price and to incentivize clients to adopt certain power consumption habits. The second objective function is presented to minimize the peak load while considering electricity costs with the DR program. Because SMGW has low load capacity, the maintenance costs of the overall system can become high within a few hours of the peak load. Therefore, reducing the peak load by several hours can help reduce the maintenance cost of the entire system and robustly design the system. The smart metering system can schedule shiftable loads depending on the proposed power management. However, non-shiftable and interruptible loads cannot be scheduled, and power must be supplied for these loads immediately when required.

$$\min_{\substack{O_{shiftable,1}(t),\ldots, \\ O_{shiftable,I}(t), \\ O_{non-shiftable,1}(t),\ldots, \\ O_{non-shiftable,J}(t), \\ O_{interuptible,1}(t),\ldots, \\ O_{interuptible,K}(t), \\ PW_{PV}(t), O_{PV}(t)}} \begin{bmatrix} \sum_{i=1}^{I} PW_{shiftable,i}(t) * O_{shiftable,i}(t) \\ +\sum_{j=1}^{J} PW_{non-shiftable,j}(t) * O_{non-shiftable,j}(t) \\ +\sum_{k=1}^{K} PW_{interruptible,k}(t) * O_{interruptible,k}(t) \\ -PW_{PV}(t) * O_{PV}(t) \end{bmatrix} * DR(t)$$

(10)

where $PW_{shiftable,i}(t)$ is the power consumption for $i$-th shiftable load at time $t$, $O_{shiftable,i}(t)$ is the switch function for $i$-th shiftable load at time $t$, $PW_{non-shiftable,j}(t)$ is the power consumption for $j$-th non-shiftable load at time $t$, $O_{non-shiftable,j}(t)$ is the switch function for $j$-th non-shiftable load at time $t$, $PW_{interruptible,k}(t)$ is the power consumption for $k$-th interruptible load at time $t$, and $O_{interruptible,k}(t)$ is the switch function for $k$-th interruptible load in energy resources at time $t$. $DR(t)$ describes price-based DR program at time $t$. $O_{shiftable,1}(t)$ ,…, $O_{shiftable,i}(t)$, $O_{non-shiftable,1}(t)$ ,…, $O_{non-shiftable,j}(t)$, and $O_{interruptible,1}(t)$,…, $O_{interruptible,k}(t)$ are controlled by the second objective function on the demand side.

### 5.3 | Optimal BSS/EV charging/discharging method

In power management, the charging/discharging mode of BSS/EV is determined based on the comparison of the grid power consumption and PV power consumption, as well as the state of charge (SOC) value of BSS/EV.

### 5.3.1 | Conventional BSS/EV charging/discharging method

The charging/discharging mode of BSS/EV at time $t$ is represented by the switching functions $O_{BSS}(t)$ and $O_{EV}(t)$, respectively. The owner of BSS/EV determines the desired charging time interval and the minimum SOC value of EV when charging is completed. The SM operator can decide the timing of the charging/discharging of BSS/EV to increase the overall profit and energy efficiency of the smart metering system. The values of the switching functions $O_{BSS}(t)$ and $O_{EV}(t)$ are determined by the charging/discharging conditions relevant to the grid power consumption, PV power consumption, and SOC of BSS/EV, given as follows:

$$O_{BSS}(t), O_{EV}(t) = \begin{cases} -1 (discharging) & If(PW_{grid}(t) - PW_{PV}(t) > PW^{max}) \\ & and\ (SOC(t) > SOC_{min}) \\ +1 (charging) & If(PW_{grid}(t) - PW_{PV}(t) < PW^{max}) \\ & and\ (SOC(t) < SOC_{max}) \\ 0 (idle) & otherwise \end{cases}$$

(11)

where $PW^{max}$ is the maximum limit of power consumption, $SOC(t)$ is the SOC of BSS/EV at time $t$, $SOC_{min}$ is the minimum limit SOC of BSS/EV, and $SOC_{max}$ is the maximum limit SOC of BSS/EV. For BSS/EV, the operational mode can be indicated by -1 (discharging), +1 (charging), and 0 (idle). In the conventional BSS/EV charging/discharging method, if the difference between grid power consumption and PV power consumption is larger than $PW^{max}$ and the SOC values of BSS/EV are higher than $SOC_{min}$, the discharging mode is assigned to BSS/EV. If the difference between grid power consumption and PV power consumption is less than $PW^{max}$ and the SOC values of BSS/EV are lower than $SOC_{max}$, the charging mode is assigned to BSS/EV. Otherwise, idle mode is assigned to BSS/EV.

## 5.3.2 | Proposed BSS/EV charging/discharging method

When considering the proposed power management, anomaly detection is used for the stable charging/discharging of BSS/EV. Furthermore, power management can be efficiently and robustly achieved by finding an abnormal state of the predicted future values, as follows:

$$O_{BSS}(t), O_{EV}(n) = \begin{cases} -1 (discharging) & If\left(\left(\max(PW_{grid}(t) * O_{LSTM}(t), \widehat{PW}_{grid}(t))\right) - PW_{PV}(t)\right) > PW^{\max} \\ & and \ (SOC(t) > SOC_{\min}) \\ +1 (charging) & If\left(\left(\max(PW_{grid}(t) * O_{LSTM}(t), \widehat{PW}_{grid}(t))\right) - PW_{PV}(t)\right) < PW^{\max} \\ & and \ (SOC(t) < SOC_{\max}) \\ 0 (idle) & Otherwise \end{cases} \quad (12)$$

where $PW_{grid}(t)$ is the grid power consumption at time $t$, $\widehat{PW}_{grid}(t)$ is the reconstructed input value of the grid power consumption of GCN-BiLSTM autoencoder at time $t$, $PW_{PV}(t)$ is the PV power consumption at time $t$, and $O_{LSTM}(t)$ is the abnormal detection parameter at time $t$. In the proposed BSS/EV charging/discharging method, the operational mode of BSS/EV, which is composed of -1 (charging), +1 (discharging), and 0 (idle), is determined by comparing the difference in the grid power consumption and PV power consumption, including the abnormal detection parameter and the maximum limit of power consumption. If an outlier or a missing value is not detected in the process of measuring the grid power consumption through the GCN-BiLSTM autoencoder, the conventional BSS/EV charging/discharging method can be applied as mentioned in equation (11). However, if an outlier or a missing value is detected in the process of measuring the grid power consumption with the GCN-BiLSTM autoencoder, the value can be replaced by $\widehat{PW}_{grid}(t)$, as stated in equation (12) to determine the charging/discharging mode of BSS/EV. The charging/discharging condition relevant to the SOC value is applied similarly to the conventional method. In the proposed BSS/EV charging/discharging method, two functions, outlier detection and missing value imputation, are added to establish a secure and robust BSS/EV charging/discharging method.

# 6 | PERFORMANCE EVALUATION

## 6.1 | DR program for peak, progressive, and climate

DR program is composed of three types of electricity rates, including time of use (ToU) based electricity rate, progressive-based electricity rate, and climate change and environmental charge rate, and they are shown as:

$$DR = (R_{TOU}, R_{PROG}, R_{CCEC}) \quad (13)$$

$$R_{TOU} = \begin{cases} 0.06\,\$/kWh\,(off-peak) & If(23:00 < t \le 09:00) \\ 0.12\,\$/kWh\,(mid-peak) & If\begin{pmatrix}09:00 < t \le 10:00\\12:00 < t \le 13:00\\17:00 < t \le 23:00\end{pmatrix} \\ 0.18\,\$/kWh\,(on-peak) & If\begin{pmatrix}10:00 < t \le 12:00\\13:00 < t \le 17:00\end{pmatrix} \end{cases} \quad (14)$$

$$R_{PROG} = \begin{cases} 0.008\,\$/kWh & If(PW_{total} \le 300kWh) \\ 0.018\,\$/kWh & If(301kWh < PW_{total} \le 450kWh) \\ 0.027\,\$/kWh & If(450kWh < PW_{total}) \end{cases} \quad (15)$$

$$R_{CCEC} = R_{RPS} + R_{ETS} + R_{CGR} \quad (16)$$

$$R_{TOU} = \begin{cases} 0.06\,\$/kWh\,(off-peak) & If(23:00 < t \le 09:00) \\ 0.12\,\$/kWh\,(mid-peak) & If\begin{pmatrix}09:00 < t \le 10:00\\12:00 < t \le 13:00\\17:00 < t \le 23:00\end{pmatrix} \\ 0.18\,\$/kWh\,(on-peak) & If\begin{pmatrix}10:00 < t \le 12:00\\13:00 < t \le 17:00\end{pmatrix} \end{cases} \quad (14)$$

$$R_{PROG} = \begin{cases} 0.008\,\$/kWh & If(PW_{total} \le 300kWh) \\ 0.018\,\$/kWh & If(301kWh < PW_{total} \le 450kWh) \\ 0.027\,\$/kWh & If(450kWh < PW_{total}) \end{cases} \quad (15)$$

$$R_{CCEC} = R_{RPS} + R_{ETS} + R_{CGR} \quad (16)$$

where $R_{TOU}$ is the ToU-based electricity rate, $R_{PROG}$ is the progressive-based electricity rate, $R_{CCEC}$ is the climate change and environmental charge rate, $R_{RPS}$ is the renewable portfolio standard electricity rate, $R_{ETS}$ is the emission trading system electricity rate, and $R_{CGR}$ is the coal generation reduction electricity rate. $PW_{total}$ is the total power consumption, which is calculated from the sum of the power consumption for shiftable, non-shiftable, and interruptible loads. $R_{TOU}$ varies depending on off-peak, mid-peak, and on-peak time intervals mentioned in equation (14). $R_{PROG}$ is determined, depending on the amount of $PW_{total}$ stated in equation (15), and $R_{CCEC}$ is calculated from the sum of $R_{RPS}$, $R_{ETS}$, and $R_{CGR}$ introduced in equation (16).

## 6.2 | Training setup and results for autoencoder

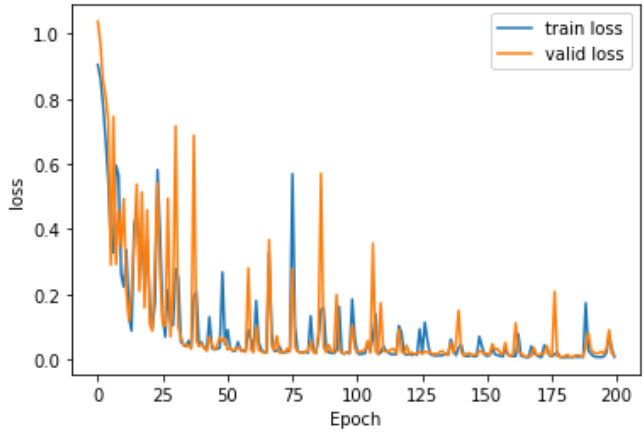

**FIGURE 5** Training of GCN-BiLSTM autoencoder.

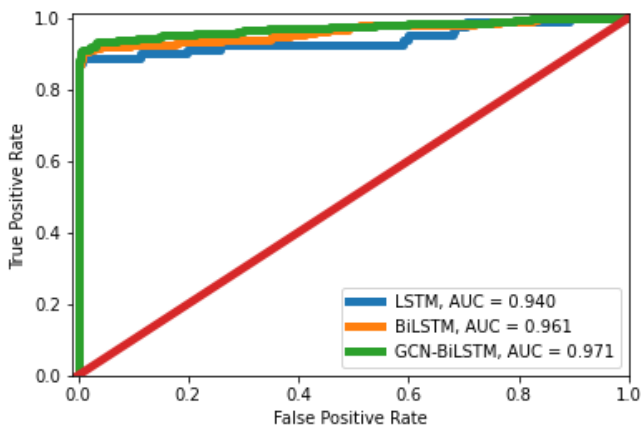

**FIGURE 6** Performance validation using receiver operating characteristic (ROC) curve for three autoencoders with load dataset.

In unsupervised learning of the GCN-BiLSTM autoencoder, the anomaly detection model is trained with only a normal dataset of gas, electric, water, heating, and hot water, which the SMGW collects from 900 households, except for outliers and missing values. Outliers and missing values are detected by the local outlier factor and K-nearest neighbor algorithm, respectively. Only normal data are used in the training of the GCN-BiLSTM autoencoder. Normal and abnormal data are used for performance validation, and the analysis of the area under the curve (AUC) – receiver operating characteristic (ROC) curve for GCN-BiLSTM, BiLSTM, and LSTM autoencoders with load dataset shown in Fig. 6 in Section 5.1.2 is performed. The abnormal dataset, which accounts for 0.8% of the entire dataset, is included in the simulation of power management with anomaly detection. The training parameters of the GCN-BiLSTM autoencoder are as follows: The ADAM optimization algorithm is used with a learning rate of 0.001, a total number of 200 training epochs, and a batch size of 128. The performance of the model with a testing dataset was often worse when trained with large batches compared to the performance obtained from small batches with spikes [52]. Small batches with spikes are typically used in training to get better performance despite fluctuations during training [53]. The loss function represents the MSE. Inadequate learning rates may result in a local minimum and overfitting. Dropout and gradients are used to avoid the local minima. The training of the GCN-BiLSTM autoencoder is performed by randomly splitting the entire dataset into 80% of the training dataset and 20% of the validation dataset, as shown in Fig. 5. The GCN-BiLSTM autoencoder has five types of input features (gas, electric, water, heating, and hot water). In Table 1, accuracy, precision, recall, F1, and MSE are illustrated, using the confusion matrix, which consists of accuracy = (TP + TN) / (TP + FN + FP + TN), precision = TP / (TP + FP), recall = TP / (TP + FN), and F1 = 2 * precision * recall / (precision + recall). TP, TN, FP, and FN are four classification results, denoting true positive, true negative, false positive, and false negative, respectively. The AUC-ROC curve, a widely used performance measure for binary classification problems, is considered in this study. The AUC values for GCN-BiLSTM, BiLSTM, and LSTM autoencoders, shown in Fig. 6, are 0.971, 0.961 and 0.94, respectively.

## 6.3 | Simulations of power management

In the simulation, it is assumed that various electrical appliances can be controlled with multi-objective optimization, and they operate with five types of integrated energy resources. Power management is performed by scheduling the controllable electric appliances through multi-objective optimization in the smart metering system, simultaneously considering the electricity consumption dataset of five types of energy resources. Scenario 1 is power management for baseline with multi-objective optimization and the conventional BSS/EV charging/discharging method without an autoencoder. Scenarios 2, 3, and 4 are the proposed power management with conventional LSTM, BiLSTM, and GCN-BiLSTM autoencoder, respectively. The gas, electric, water, heating, and hot water data, except the outliers obtained from anomaly detection, are described as an interruptible load in the power management.

In Fig. 7(a), the power consumption of shiftable, non-shiftable, and interruptible loads measured by the smart metering system shows the variation of the average power consumption of the 900 households based on four scenarios. Scenarios 1, 2, 3, and 4 are indicated by a solid blue line, solid red line, dotted sky blue line with a circle marker, and dotted black line with a triangle marker, respectively. Scenario 1 leads to the highest peak load of 3.7 kW, as shown in Table 1. Because renewable energy is abundant and the charging/discharging of BSS/EV is actively achieved in the afternoon, the peak load occurs around 20:00. In scenarios 2, 3, and 4, charging/discharging of BSS/EV with the autoencoder is more active than Scenario 1, resulting in reduced grid power consumption. Furthermore, it affects the reduction of peak load and electricity costs. Scenario 4 demonstrates better power management than scenarios 1, 2, and 3 because the autoencoder uses a more accurate GCN-BiLSTM network than the conventional LSTM and BiLSTM networks.

Figure 7(b) depicts the average BSS/EV power consumption of 900 households based on the four scenarios. The charging rate of BSS/EV is 1–3 kW, and the charging/discharging operation of BSS/EV is activated in the order of scenarios 4, 3, 2, and 1. The HVAC power consumption of the smart metering system is presented in Fig. 7(c). Since all four scenarios assume the same HVAC operating conditions, the power consumption of HVAC is identical, regardless of the scenario. Fig. 7(d) depicts the power consumption of the electric appliances measured by the smart metering system, which shows similar power consumption levels within the four scenarios, as shiftable loads are similarly scheduled depending on other load demands.

Table 1 lists the simulation results based on scenarios 1, 2, 3, and 4. The comparison of performance measurements of the autoencoders for scenarios 2, 3, and 4 are conducted based on accuracy, precision, recall, $F_1$, and MSE. The results indicate that the performance of the GCN-BiLSTM autoencoder is better than that of the conventional LSTM autoencoder and Bi-LSTM autoencoder. Notably, the GCN-BiLSTM autoencoder used in Scenario 4 produces the best results in terms of the reduction of peak load and electricity costs with the smart metering system.

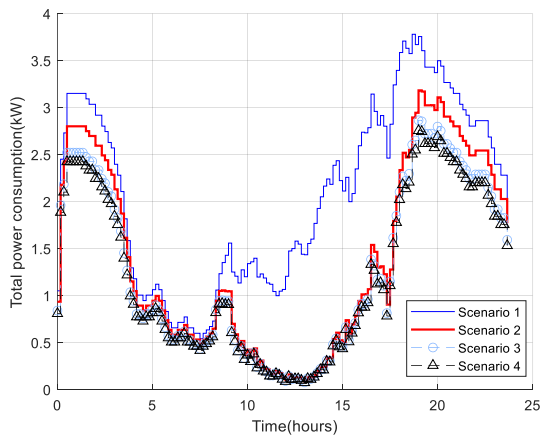

(a)

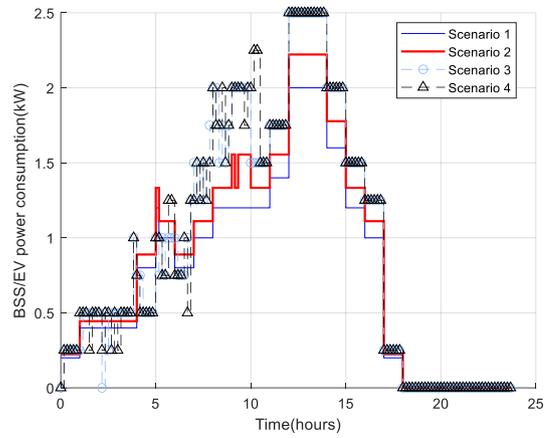

(b)

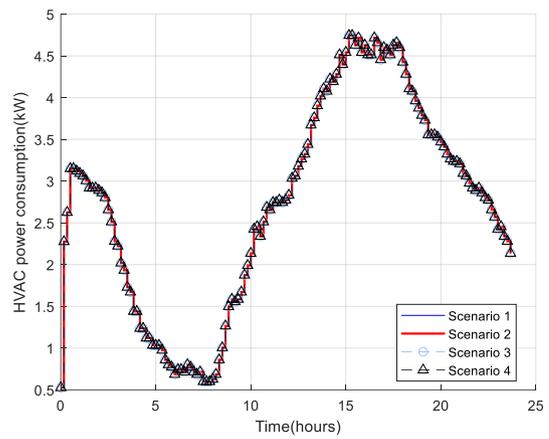

(c)

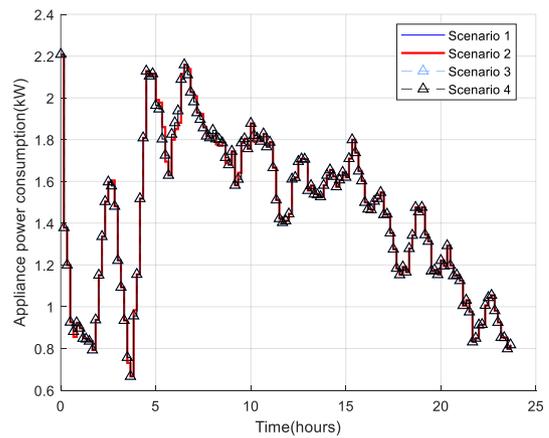

(d)

**FIGURE 7** Simulation results measured by smart metering system: (a) Total power consumption for shiftable, non-shiftable, and interruptible loads; (b) BSS/EV power consumption; (c) HVAC power consumption; (d) Electric appliances power consumption.

**TABLE 1** Simulation results

| Category | Accuracy | Precision | Recall | $F_1$ | MSE | Peak Load | Electricity Cost |
|---|---|---|---|---|---|---|---|
| Scenario 1 | - | - | - | - | - | 3.7kW | $44.8 |
| SCENARIO 2 | 0.9925 | 0.9997 | 0.99976 | 0.99626 | 0.34 | 3.1 kW | $31.7 |
| SCENARIO 3 | 0.99336 | 0.99348 | 0.99984 | 0.99665 | 0.32 | 2.8 kW | $30.5 |
| SCENARIO 4 | 0.99293 | 0.99302 | 0.99988 | 0.99644 | 0.31 | 2.8 kW | $30.3 |

# 7 | CONCLUSION

Power management with anomaly detection achieved by a novel smart metering system was proposed in this study. The proposed scheme is featured by AMI, SM data processing procedure, smart metering cloud system design, and five types of energy resources: gas, electric, water, heating, and hot water. SM data was obtained from 900 households. An autoencoder with a GCN-BiLSTM network was used to find outliers and missing values for anomaly detection of five types of energy resources, and its prediction accuracy was evaluated. The proposed power management was implemented, considering charging/discharging of BSS/EV. For power management, multi-objective optimization was performed with an objective function to maximize the profit of REPs and minimize the grid dependency and another function to minimize the peak load while considering the DR program. In the simulations, the performance of the autoencoders was evaluated in terms of accuracy, precision, recall, F1, and MSE. The autoencoder using the GCN-BiLSTM network enabled more reduced peak load and electricity costs through more accurate anomaly detection. The proposed power management with the functionality of anomaly detection in the smart metering system reduced the three types of tariff-based electricity costs and the amount of power supplied from the grid.

## ACKNOWLEDGMENTS

This work was supported by the Korea Institute of Energy Technology Evaluation and Planning and the Ministry of Trade, Industry & Energy of the Republic of Korea (No. 20191210301580, No. 20192010107290).